\documentclass[aip,reprint,graphicx]{revtex4-2}
\usepackage{multirow}
\usepackage{mathtools}
\usepackage{txfonts}
\usepackage{bm}
\DeclareUnicodeCharacter{2009}{\,} 
\usepackage[dvipsnames]{xcolor}
\usepackage{array}

\begin{document}

\title{Stochastic evaluation of  four-component relativistic second-order many-body perturbation energies: A potentially quadratic-scaling correlation method}

\author{J. C\'{e}sar Cruz}
\affiliation{Departamento de Qu\'imica, Divisi\'on de Ciencias B\'asicas e Ingenier\'ia, Universidad Aut\'onoma Metropolitana-Iztapalapa, San Rafael Atlixco 186, Col.~Vicentina, Iztapalapa, C.~P.~09340, Ciudad de M\'exico, M\'exico}
\author{Jorge Garza}
\affiliation{Departamento de Qu\'imica, Divisi\'on de Ciencias B\'asicas e Ingenier\'ia, Universidad Aut\'onoma Metropolitana-Iztapalapa, San Rafael Atlixco 186, Col.~Vicentina, Iztapalapa, C.~P.~09340, Ciudad de M\'exico, M\'exico}
\author{Takeshi Yanai}
\affiliation{Department of Chemistry, Graduate School of Science, Nagoya University, Furocho, Chikusa Ward, Nagoya, Aichi 464-8601, Japan}
\affiliation{Institute of Transformative Bio-Molecules (WPI-ITbM), Nagoya University, Furocho, Chikusa Ward, Nagoya, Aichi 464-8601, Japan}
\author{So Hirata}
\affiliation{Department of Chemistry, University of Illinois at Urbana-Champaign, 600 South Mathews Avenue, Urbana, Illinois 61801, USA}
\email[]{sohirata@illinois.edu}

\date{\today}

\begin{abstract}
A second-order many-body perturbation correction to the relativistic Dirac--Hartree--Fock energy is evaluated stochastically
by integrating 13-dimensional products of four-component spinors and Coulomb potentials. 
The integration in the real space of electron coordinates is carried out
by the Monte Carlo (MC) method with the Metropolis sampling, whereas the MC integration in the imaginary-time domain is performed by 
the inverse-CDF (cumulative distribution function) method. 
The computational cost to reach a given relative statistical error for spatially compact but heavy molecules  
is observed to be no worse than cubic and possibly quadratic with the number of electrons or basis functions.  
This is a vast improvement over the quintic scaling of
the conventional, deterministic second-order many-body perturbation method.
The algorithm is also
easily and efficiently parallelized with demonstrated 92\% strong scalability going from 64 to 4096 processors.
\end{abstract}

\maketitle 

\section{Introduction}

The algorithms of {\it ab initio} electron-correlation methods are dominated by matrix multiplications.
Implemented naively, such multiplications are often nonscalable with respect to either computer size or system size. 
The scalability with respect to computer size means the extent to which an algorithm lends itself to a facile and efficient 
parallel execution on a large number of computing processors, which is, nowadays, on the order of millions. 
The scalability with system size concerns with the size-dependence of the computational cost,
which ranges from $O(n^5)$ for the second-order many-body perturbation (MP2) method \cite{Moller_1934} to an exponential function of $n$ for the full configuration-interaction method,\cite{ Knowles_1984, Sherrill_1999}
where $n$ is a measure of system size such as the number of electrons or of basis functions.

Consider the four-index integral transformation.\cite{Yoshimine_1973, Abe_2004_1} This is a necessary step for most electron-correlation calculations,
and is the rate-determining step of MP2.\cite{headgordon1988, frisch1990,frisch1990_2}
In it, two-electron integrals in the atomic-orbital (AO) basis are transformed to those in the molecular-orbital (MO) basis
via four consecutive quarter transformations (which are binary matrix multiplications). This is a hard step to program and even harder to parallelize for the following reasons:
Neither the AO-based nor MO-based two-electron integrals fit in core memory, and so they (together with the partially transformed integrals) have 
to be stored on disks at a large I/O cost. The AOs and MOs have opposing characteristics---AOs are spatially local but non-symmetry-adapted, while
MOs are spatially extended and symmetry-adapted---hindering their data-local storage across processors, 
which is prerequisite for efficient parallelization. Each of the four quarter transformations comes at a $O(n^5)$ cost, raising a high barrier 
for {\it ab initio} electron-correlation calculations, as compared to preceding reference mean-field calculations, which can be completed at a $O(n^2)$ to $O(n^3)$ cost.

There are at least two general approaches to combating the nonscalability: One is to exploit the sparsity of 
the matrices, such as that brought to by spatially localizing MOs.\cite{Pulay1,Yang_1995,Kohn} This will divide a large matrix multiplication
into many smaller ones, each associated with a local domain, potentially lowering size-dependence of cost all the way down to $O(n^1)$. 
Each of the small matrix multiplications can be taken care of by an individual processor with minimal interprocessor communications,
exposing efficient parallelism. The price one pays for this remarkable scalability is its limited applicability 
only to {\it spatially extended large systems} (additionally, to those that have local electronic structures, barring, e.g., metals and charge transfer).

The other approach is to not exhaustively use all matrix elements during the multiplication in favor of stochastically sampling
only a small portion of them according to their importance.\cite{Thom2007, Ohtsuka2008, Booth2009, Cleland2010, Ohtsuka2010, Ohtsuka2011, Shepherd2012, Petruzielo2012, Willow2012, Neuhauser2013, Neuhauser2013b, Willow2013, Ten-no2013, Willow2013b, Booth2014, Cytter2014, Ge2014, Willow2014, Willow2014a, Willow2014b, Doran2016, Johnson2016, piecuch,Neuhauser2017, Takeshita2017, Jeanmairet2017, Garniron2017, Ten-no2017, Johnson2018,Spencer2018,  Doran2019, Filip2019, Dou2019, Caffareld2019, Li2019, Doran2020a, Doran2020b,DoranQiu} For instance, the Monte Carlo MP2 (MC-MP2) method\cite{Willow2012,Willow2013b,Doran2016} recasts the MP2 energy formula
into the sum of two 13-dimensional integrals, which are evaluated by the MC integration method, eradicating the four-index integral computation, transformation, or storage. 
Each processor can be tasked with a completely independent
MC integration in a parallel execution. Only the final results from surviving processors need to be averaged 
with near perfect parallel efficiency on
up to thousands of CPUs (central processing units) or on hundreds of GPUs (graphical processing units).\cite{Doran2016}
Size-dependence of cost is not known {\it a priori}, as it depends on the suitability of weight functions adopted for importance sampling in 
the high-dimensional space. Our numerical observation\cite{Doran2016} indicated that it is $O(n^3)$, two ranks lower than the deterministic MP2's $O(n^5)$ cost.\cite{headgordon1988, frisch1990,frisch1990_2} 
Note that some incarnations of quantum Monte Carlo (QMC)\cite{Ceperley:1980uz, Luchow:2000ji, Foulkes, Kolorenc:2011hv, AustinLester, Wagner,Toulouse2016, Kunitsa}  solve the Schr\"odinger equation essentially exactly also at a $O(n^3)$ cost.
The tradeoff for this greater scalability is the introduction of statistical errors.\cite{Ceperley_error}

It is of interest to study whether the reduced scaling of stochastic algorithms still holds for {\it spatially compact  large systems} such as heavy-element compounds. 
They have a large number of electrons and basis functions, but in a small volume, and therefore the local-correlation approach would not work. 
Stochastic algorithms such as QMC including MC-MP2 could be among the few that can potentially reduce size-dependence 
of cost in such crowded systems. For heavy-element compounds, it is imperative to include the effects of special theory of relativity.\cite{Yanai_2004,Nakajima_2005,Liu_2010, Wullen_2012,Liu_2020} A relativistic MP2 theory
based on the Dirac--Hartree--Fock reference is well established,\cite{Dyall_1994, Laerdahl_1997, Laerdahl_1998, Abe_2004,Abe_2005} and its conventional 
algorithmic hotspot is still the four-index integral transformation,\cite{Abe_2004_1, Abe_2006} which  
is a $O(n^5)$ procedure. The objective of this article is to demonstrate numerically that the Monte Carlo algorithm of the four-component relativistic MP2 
method
preserves the reduced scaling for such spatially compact, crowded systems by eradicating the integral transformation step. 
Numerical evidence obtained in this study suggests that the scaling may be as low as 
$O(n^2)$. 

\section{Four-component relativistic MP2 theory}

Here, we succinctly review only the salient aspects of the four-component relativistic quantum chemistry\cite{Grant_1979, Grant_1988} as a basis of the relativistic MC-MP2 algorithm described
in the next section. The Monte Carlo algorithm is not limited to the four-component relativistic approach and should be equally valid for 
the two-component (A2C),\cite{ZORA_1995, Lenthe_ZORA} exact two-component (X2C),\cite{Liu_X2C, Saue_X2C} and quasi-four-component (Q4C)\cite{Liu_Q4C, Peng_Q4C} 
approaches not to mention the one-component and nonrelativistic ones.\cite{Willow2012, Willow2013, Willow2013b}

The Dirac--Coulomb(--Breit) Hamiltonian\cite{Dyall_book} for a molecule with $n_\text{elec}$ electrons and $n_\text{nuc}$ nuclei is written as
\begin{eqnarray}
\hat{H} = \sum_i^{n_\text{elec}} \hat{h}_\text{D}(i) +  \sum_{i < j}^{n_\text{elec}} \hat{g}_{ij} + \sum_{I < J}^{n_\text{nuc}} \frac{Z_I Z_J}{|\bm{R}_I-\bm{R}_J|}, 
\end{eqnarray}
where $Z_I$ and $\bm{R}_I$ are the charge and position, respectively, of the $I$th nucleus, and $\hat{g}_{ij}$ is the electron-electron interaction operator defined later. 
Here, $\hat{h}_\text{D}(i)$ is the one-electron Dirac Hamiltonian for the $i$th electron given by
\begin{equation}
\hat{h}_\text{D}(i) = c \bm{\alpha} \cdot \bm{p}(i) + \beta  c^2 + V_\text{nuc}(i),
\end{equation}
where $c$ is the speed of light, $\bm{p}(i)$ is the momentum operator that is equal to $-i (\partial /\partial x_i, \partial /\partial y_i, \partial /\partial z_i)$, 
and $V_\text{nuc}(i)$ is the nuclear-attraction potential defined later, while
$\bm{\alpha}=(\alpha_1,\alpha_2,\alpha_3)$ and ${\beta}$ are $4\times4$ matrices of the forms,
\begin{eqnarray}
{\alpha}_k &=& 
\left( \begin{array}{cc} 
\bm{0}_2 & \bm{\sigma}_k \\ 
\bm{\sigma}_k & \bm{0}_2 \\ 
\end{array} \right), \,\,\, (k = 1, 2, 3), \label{alpha} \\
{\beta} &=& \left( \begin{array}{cc} 
\bm{1}_2 & \bm{0}_2 \\ 
\bm{0}_2 & -\bm{1}_2 \\ 
\end{array} \right), 
\end{eqnarray}
where $\bm{1}_2$ and $\bm{0}_2$ are the $2 \times 2$ unit and null matrices, respectively, and $\bm{\sigma}_k$ ($k= 1, 2, 3$) are the Pauli matrices.\cite{Dyall_book}

The nuclear-attraction potential $V_\text{nuc}(i)$ in the point-nucleus approximation is given by
\begin{eqnarray}
V_\text{nuc}(i) = - \sum_{I}^{n_\text{nuc}} \frac{Z_I}{|\bm{r}_i - \bm{R}_I|}.
\end{eqnarray}
The interaction operator $\hat{g}_{ij}$ between
the $i$th and $j$th electrons is written as
\begin{equation}
\hat{g}_{ij} = \frac{1}{r_{ij}} -  \frac{\bm\alpha(i) \cdot \bm\alpha(j)}{2r_{ij}} - \frac{(\bm\alpha(i)\cdot \bm{r}_{ij}) (\bm\alpha(j) \cdot \bm{r}_{ij})}{2 r_{ij}^3},
\end{equation}
where $\bm{r}_{ij} = \bm{r}_i - \bm{r}_j$, $r_{ij} = |\bm{r}_{ij}|$, and
$\bm{\alpha}(i)$ is the $\bm\alpha$ matrix-vector of Eq.\ (\ref{alpha}) acting on the $i$th electron.
The first term is the Coulomb interaction, while the other two terms are the Breit interaction.\cite{Liu_2010, Liu_2014} In this work, only the Coulomb interaction
was retained with the Breit interaction neglected. As it will become evident in the next section, however, including the Breit interaction
will be straightforward in the MC-MP$n$ methods because the associated integrals are evaluated numerically. Likewise,
the use of finite nucleus models\cite{Abe_2008} should also be much easier in the MC algorithm than in the deterministic ones. These may be considered in a future study.  

An eigenfunction (spinor) of $\hat{h}_\text{D}(i)$ is a four-component complex vector,
\begin{eqnarray}
\varphi_k = 
\left( \begin{array}{c}
\varphi_{k}^{L\alpha} \\
\varphi_{k}^{L\beta} \\
\varphi_{k}^{S\alpha} \\
\varphi_{k}^{S\beta} \\
\end{array}
\right),
\end{eqnarray}
where $L$ and $S$ stand for the large and small components of the spinor, respectively, while $\alpha$ and $\beta$ signify spin. 
Eigenvalues are divided into two groups: One with positive energies ($E > mc^2$) and the other with negative energies ($E < -mc^2$). 
The positive energies correspond to electrons with large $L$ amplitudes, while the negative energies to positrons with large $S$ amplitudes.

A wave function of the mean-field (Dirac--Hartree--Fock) theory is the Slater determinant of positive-energy spinors occupied by electrons, which minimizes
its expectation value of the Dirac--Coulomb(--Breit) Hamiltonian. The spinors are expanded in a four-component basis set as
\begin{eqnarray}
\varphi_k = \sum_{\mu}^{n_\text{bas}} 
\left( \begin{array}{c}
C_{\mu k}^{L}\, \chi_{\mu}^{L\alpha} \\
C_{\mu k}^{L}\, \chi_{\mu}^{L\beta} \\
C_{\mu k}^{S}\, \chi_{\mu}^{S\alpha} \\
C_{\mu k}^{S}\, \chi_{\mu}^{S\beta} \\
\end{array}
\right). \label{C}
\end{eqnarray}
Each component basis function is a product of a radial Gaussian function and spherical harmonic, although the MC algorithm can use any basis function.
The $L$ and $S$ components of a basis function 
have to satisfy the kinetic balance condition,\cite{Ishikawa_1984, Nakajima_2005}
\begin{equation}
\left( \begin{array}{c} 
\chi_{\mu}^{S\alpha} \\
\chi_{\mu}^{S\beta} \\
\end{array} \right)
= i (\bm\sigma\cdot \bm{p})\
\left( \begin{array}{c} 
\chi_{\mu}^{L\alpha} \\
\chi_{\mu}^{L\beta} \\
\end{array} \right),
\end{equation} 
to avoid a variational collapse. This causes the $S$-component basis set to be about twice as large as the $L$-component one, although an innovation
was introduced by one of the authors that keeps the sizes of these basis sets equal.\cite{Yanai_2001-114}
In either case, the number of basis functions is typically two to three times greater 
in a four-component relativistic calculation than in the corresponding nonrelativistic one, making the integral evaluation and subsequent transformation arduous.\cite{Abe_2004_1} Time-reversal symmetry renders each eigenvalue at least doubly degenerate for an even number of electrons so that the corresponding two spinors 
form a Kramers pair.\cite{Dyall_book} This along with spatial symmetry (double group symmetry)\cite{Dyall_book} can be exploited 
to reduce the number of unique integrals and thus its computational cost in a deterministic algorithm. The exploitation of 
symmetry, however, seems incompatible with the MC algorithm. 

The zeroth-order M{\o}ller--Plesset perturbation (MP0) energy is the sum of occupied spinor energies, which double counts
the electron-electron interaction energy and is thus too negative.
The sum of the MP0 and first-order (MP1) energies recovers the Dirac--Hartree--Fock energy. The second-order correction, therefore,
accounts for the electron-correlation effect for the first time, and is given by\cite{Dyall_1994, Dyall_book, Laerdahl_1997}
\begin{eqnarray}\label{FinalMP2}
E^{(2)} = \frac{1}{2}\sum_{i,j}^{\text{occ.}} \sum_{a,b}^{\text{vir.}} \frac{(ia|jb)[(ai|bj) - (bi|aj)] }{ \epsilon_i + \epsilon_j - \epsilon_a - \epsilon_b} ,
\end{eqnarray}
where $i$ and $j$ run over all occupied positive-energy spinors, while $a$ and $b$ over all virtual positive-energy spinors, and $\epsilon_p$
is the $p$th spinor energy. That the negative-energy spinors do not enter the summations is called the no-pair approximation.\cite{Sucher_1980, Auts_2012, Liu_2014}
For an even number of electrons, this formula can be compressed considerably by using time-reversal symmetry and turned into the so-called
Kramers-restricted formula.\cite{Dyall_1994, Laerdahl_1997, Laerdahl_1998} We do not consider this here because, again, the MC algorithm seems incompatible with symmetry advantages. 
There are other fast MP2 methods such as the resolution-of-identity\cite{RIMP2} and local approximations,\cite{localMP2} which should be compatible with and complementary to the MC algorithm.

In Eq.\ (\ref{FinalMP2}), $(ia|jb)$ is a two-electron integral in the Dirac--Hartree--Fock spinor basis, defined by\cite{Almou_2016}
\begin{eqnarray}
(ia|jb) = \iint d\bm{r}_1 d\bm{r}_2\, { \varphi_i^{\dagger}(\bm{r}_1)\varphi_a(\bm{r}_1) \hat{g}_{12}\,  \varphi_j^{\dagger}(\bm{r}_2) \varphi_b(\bm{r}_2)},
\end{eqnarray}
which is related to the same in the AO basis functions via MO coefficients $\{ C^L_{\zeta p} \}$ and $\{ C^S_{\zeta p} \}$ as
\begin{eqnarray}
(ia|jb) &=& \sum_{\mu}^{n_\text{bas}} \sum_{\nu}^{n_\text{bas}} \sum_{\kappa}^{n_\text{bas}} \sum_{\lambda}^{n_\text{bas}} 
C^{L*}_{\mu i} C^L_{\nu a} C^{L*}_{\kappa j} C^L_{\lambda b} (\mu^{L\alpha} \nu^{L\alpha}|\kappa^{L\alpha}\lambda^{L\alpha}) 
\nonumber\\ && + 
\sum_{\mu}^{n_\text{bas}} \sum_{\nu}^{n_\text{bas}} \sum_{\kappa}^{n_\text{bas}} \sum_{\lambda}^{n_\text{bas}} 
C^{L*}_{\mu i} C^L_{\nu a} C^{L*}_{\kappa j} C^L_{\lambda b} (\mu^{L\alpha} \nu^{L\alpha}|\kappa^{L\beta}\lambda^{L\beta}) 
\nonumber\\ && + 
\sum_{\mu}^{n_\text{bas}} \sum_{\nu}^{n_\text{bas}} \sum_{\kappa}^{n_\text{bas}} \sum_{\lambda}^{n_\text{bas}} 
C^{L*}_{\mu i} C^L_{\nu a} C^{S*}_{\kappa j} C^S_{\lambda b} (\mu^{L\alpha} \nu^{L\alpha}|\kappa^{S\beta}\lambda^{S\beta}) 
\nonumber\\ && + \dots \,\,\, (\text{12 terms})
\nonumber\\ && + 
\sum_{\mu}^{n_\text{bas}} \sum_{\nu}^{n_\text{bas}} \sum_{\kappa}^{n_\text{bas}} \sum_{\lambda}^{n_\text{bas}} 
C^{S*}_{\mu i} C^S_{\nu a} C^{S*}_{\kappa j} C^S_{\lambda b} (\mu^{S\beta} \nu^{S\beta}|\kappa^{S\beta}\lambda^{S\beta}) \nonumber\\
\label{transformation}
\end{eqnarray}
with
\begin{eqnarray}
&&(\mu^{L\alpha}\nu^{L\alpha}|\kappa^{L\beta}\lambda^{L\beta}) \nonumber\\
&&= \iint d\bm{r}_1 d\bm{r}_2\, { \chi_\mu^{L\alpha*}(\bm{r}_1)\chi_\nu^{L\alpha}(\bm{r}_1) \hat{g}_{12}\,  \chi_\kappa^{L\beta*}(\bm{r}_2) \chi_\lambda^{L\beta}(\bm{r}_2)},
\end{eqnarray}
etc., although the actual implementation of this step is more compact and efficient than these equations may imply.\cite{Abe_2004_1}
Nonetheless, the four-index integral transformation [Eq.\ (\ref{transformation})] is the algorithmic hotspot of the four-component relativistic MP2, having a $O(n^5)$ cost, if we assume $n \propto n_\text{elec} \propto n_\text{bas}$. In the MC algorithm, this step is altogether eliminated along with the need to store original, partially, or fully transformed integrals. This algorithm is now described. 

\section{Monte Carlo relativistic MP2 algorithm\label{sec:MC}}

Using the following Laplace transform,\cite{Amlof_1991,Haser_1992}
\begin{eqnarray}
\frac{1}{\epsilon_i + \epsilon_j - \epsilon_a - \epsilon_b} = - \int_0^{\infty} d\tau\,  \exp \left[(\epsilon_i + \epsilon_j - \epsilon_a - \epsilon_b)\tau\right],
\end{eqnarray}
we convert Eq.\ (\ref{FinalMP2}) into the sum of two 13-dimensional integrals,\cite{Willow2012} i.e.,
\begin{eqnarray}
E^{(2)} &=& -2 \iiiint d\bm{r}_1d\bm{r}_2d\bm{r}_3d\bm{r}_4 \int_0^\infty d\tau\,\frac{1}{r_{12}r_{34}} \nonumber\\
&& \times \,\left\{  \text{vec}\left[ \bm{G}^-(\bm{r}_1,\bm{r}_3,-\tau)\right]^T \cdot \text{vec} \left[ \bm{G}^+(\bm{r}_3,\bm{r}_1,\tau)\right] \right\}  
\nonumber\\&& \times\,
 \left\{ \text{vec}\left[ \bm{G}^-(\bm{r}_2,\bm{r}_4,-\tau)\right]^T  \cdot \text{vec} \left[ \bm{G}^+(\bm{r}_4,\bm{r}_2,\tau)\right]  \right\} \nonumber \\
&& + \iiiint d\bm{r}_1d\bm{r}_2d\bm{r}_3d\bm{r}_4 \int_0^\infty d\tau\,\frac{1}{r_{12}r_{34}} \nonumber\\
&& \times \,\left\{  \text{vec}\left[ \bm{G}^-(\bm{r}_1,\bm{r}_3,-\tau)\right]^T \cdot \text{vec} \left[ \bm{G}^+(\bm{r}_4,\bm{r}_1,\tau)\right] \right\}  
\nonumber\\&& \times\,
 \left\{ \text{vec}\left[ \bm{G}^-(\bm{r}_2,\bm{r}_4,-\tau)\right]^T  \cdot \text{vec} \left[ \bm{G}^+(\bm{r}_3,\bm{r}_2,\tau)\right]  \right\} , \label{MCMP2}
\end{eqnarray}
where $\tau$ has the physical meaning of imaginary time, and 
$\text{vec}[\dots]$ is the vectorization operator that 
reshapes a $m \times n$ matrix  
into a $mn \times 1$  column vector. Green's function traces, $\bm{G}^\pm(\bm{r}_d,\bm{r}_o,\tau)$, are $4 \times 4$ matrices defined by
\begin{eqnarray}
&& \bm{G}^+(\bm{r}_d,\bm{r}_o,\tau) \nonumber\\
&& = \sum_{a}^{\text{vir.}} \varphi_a(\bm{r}_d) \varphi_a^\dagger(\bm{r}_o) \exp(-\epsilon_a \tau) \nonumber\\
&& = \sum_{a}^{\text{vir.}} \left( 
\begin{array}{c}
\varphi^{L\alpha}_a(\bm{r}_d) \\
\varphi^{L\beta}_a(\bm{r}_d) \\
\varphi^{S\alpha}_a(\bm{r}_d) \\
\varphi^{S\beta}_a(\bm{r}_d) \\
\end{array}
\right) \left(
\begin{array}{cccc} 
\varphi^{L\alpha*}_a(\bm{r}_o) &
\varphi^{L\beta*}_a(\bm{r}_o) &
\varphi^{S\alpha*}_a(\bm{r}_o) &
\varphi^{S\beta*}_a(\bm{r}_o)
\end{array}
\right) 
\nonumber\\&& \times
\exp(-\epsilon_a \tau), \\
&& \bm{G}^-(\bm{r}_d,\bm{r}_o,\tau) \nonumber\\
&&= \sum_{i}^{\text{occ.}} \varphi_i(\bm{r}_d) \varphi_i^\dagger(\bm{r}_o) \exp(-\epsilon_i \tau) \nonumber\\
&&=  \sum_{i}^{\text{occ.}} \left( 
\begin{array}{c}
\varphi^{L\alpha}_i(\bm{r}_d) \\
\varphi^{L\beta}_i(\bm{r}_d) \\
\varphi^{S\alpha}_i(\bm{r}_d) \\
\varphi^{S\beta}_i(\bm{r}_d) \\
\end{array}
\right) \left(
\begin{array}{cccc} 
\varphi^{L\alpha*}_i(\bm{r}_o) &
\varphi^{L\beta*}_i(\bm{r}_o) &
\varphi^{S\alpha*}_i(\bm{r}_o) &
\varphi^{S\beta*}_i(\bm{r}_o)
\end{array}
\right) 
\nonumber\\&& \times
\exp(-\epsilon_i \tau).
\end{eqnarray}

For example, one of the factors in Eq.\ (\ref{MCMP2}) is more explicitly written as  
\begin{eqnarray}
&& \left\{  \text{vec}\left[ \bm{G}^-(\bm{r}_1,\bm{r}_3,-\tau)\right]^T \cdot \text{vec} \left[ \bm{G}^+(\bm{r}_3,\bm{r}_1,\tau)\right] \right\}  \nonumber\\
&& =  
O^{L\alpha L\alpha}(\bm{r}_1,\bm{r}_3,-\tau) V^{L\alpha L\alpha}(\bm{r}_3,\bm{r}_1,\tau) \nonumber\\&&
+ O^{L\alpha L\beta}(\bm{r}_1,\bm{r}_3,-\tau) V^{L\alpha L\beta}(\bm{r}_3,\bm{r}_1,\tau) \nonumber\\&&
+ O^{L\alpha S\alpha}(\bm{r}_1,\bm{r}_3,-\tau) V^{L\alpha S\alpha}(\bm{r}_3,\bm{r}_1,\tau) \nonumber\\&&
+\dots \,\,\, (\text{12 terms}) \nonumber\\&&
+ O^{S\beta S\beta}(\bm{r}_1,\bm{r}_3,-\tau) V^{S\beta S\beta}(\bm{r}_3,\bm{r}_1,\tau)
\end{eqnarray}
with
\begin{eqnarray}
O^{L\alpha L\beta}(\bm{r}_1,\bm{r}_3,-\tau)  &=& \sum_{i}^{\text{occ.}} \varphi^{L\alpha}_{i}(\bm{r}_1) \varphi^{L\beta*}_{i} (\bm{r}_3) \exp(\epsilon_i \tau), \\
V^{L\alpha L\beta}(\bm{r}_3,\bm{r}_1,\tau)  &=& \sum_{a}^{\text{vir.}} \varphi^{L\alpha}_{a}(\bm{r}_3) \varphi^{L\beta*}_{a} (\bm{r}_1) \exp(-\epsilon_a \tau),
\end{eqnarray}
etc. The other factors of Eq.\ (\ref{MCMP2}) and the remaining 15 $O$ and 15 $V$ functions are defined analogously.

Rewriting Eq.\ (\ref{MCMP2}) as
\begin{eqnarray}
E^{(2)} =  \iiiint d\bm{r}_1d\bm{r}_2d\bm{r}_3d\bm{r}_4  \int_0^{\infty} d\tau\, {f}(\bm{r}_1, \bm{r}_2, \bm{r}_3, \bm{r}_4, \tau),
\end{eqnarray}
where $f$ is the sum of the two integrands, we evaluate the right-hand side by the Monte Carlo (MC) method\cite{Willow2012,Doran2019} as
\begin{eqnarray}
E^{(2)} \approx \frac{1}{N} \sum_{n=1}^N \frac{{f}(\bm{r}_1^{[n]}, \bm{r}_2^{[n]}, \bm{r}_3^{[n]}, \bm{r}_4^{[n]}, \tau^{[n]})}
{w(\bm{r}_1^{[n]}, \bm{r}_2^{[n]}) w(\bm{r}_3^{[n]}, \bm{r}_4^{[n]}) w(\tau^{[n]})} \equiv I_N, 
\end{eqnarray}
where $I_N$ is the MC estimate of the integral at the $N$th MC step. 
Electron-pair coordinates at the $n$th MC step, $(\bm{r}_1^{[n]}, \bm{r}_2^{[n]})$, are 
coupled through the Coulomb potential $1/r_{12}$ and are distributed in the six-dimensional space 
randomly but according to a judiciously chosen weight function, $w(\bm{r}_1^{[n]}, \bm{r}_2^{[n]})$.
Electron-pair coordinates, $(\bm{r}_3^{[n]}, \bm{r}_4^{[n]})$, are essentially the same, whereas $w(\tau^{[n]})$ dictates a random distribution
of the imaginary-time coordinate, $\tau^{[n]}$, at the $n$th MC step.\cite{Doran2019}

We shall reuse the weight functions\cite{Willow2012,Willow2013,Doran2019} that have been found effective for the nonrelativistic case. 
The weight function for electron pairs is the integrand of a two-center Coulomb-repulsion integral
for the sum of contracted $s$-type Gaussian-type orbitals:
\begin{eqnarray}
w(\bm{r}_1^{[n]}, \bm{r}_2^{[n]}) = \frac{1}{N_g} \frac{g(\bm{r}_1)g(\bm{r}_2)}{r_{12}}
\end{eqnarray}
with
\begin{eqnarray}\label{basisSet}
g(\bm{r}) &=& \sum_{I=1}^{n_{\text{nuc.}}} \left\{c_{1}(I) N_{\zeta_{1}(I)}\exp\left({-\zeta_{1}(I) |\bm{r}-\bm{R}_I|^2}\right) \right. 
\nonumber\\ && \left. + c_{2}(I) N_{\zeta_{2}(I)}\exp\left({-\zeta_{2}(I) |\bm{r}-\bm{R}_I|^2}\right) \right\}, 
\end{eqnarray}
where $N_\zeta$ is the normalization coefficient of each primitive Gaussian-type orbital. The overall normalization factor,
\begin{eqnarray}
N_g = \iint  {d}\bm{r}_1d\bm{r}_2\, \frac{g(\bm{r}_1)g(\bm{r}_2)}{r_{12}},
\end{eqnarray}
can be evaluated analytically.\cite{Obara1986} A random distribution of the electron-pair walkers according to this weight function is achieved by the Metropolis algorithm.\cite{Metropolis}
It may be noticed that the Coulomb potentials, $1/r_{12}$ and $1/r_{34}$, in the integrand $f$, which have singularities everywhere, 
are canceled out by the same in the weight functions.

The weight function in the $\tau$ dimension is given by\cite{Doran_2019} 
\begin{eqnarray}\label{tauWeight}
w(\tau) = \lambda \exp({-\lambda \tau}),
\end{eqnarray}
which is normalized. The exponent $\lambda$ is parametrized by the energies of the highest-occupied and lowest-unoccupied molecular orbitals (HOMO and LUMO) as
\begin{eqnarray}
\lambda = 2(\epsilon_{\text{LUMO}} - \epsilon_{\text{HOMO}}).
\end{eqnarray}
A random distribution of the imaginary-time walker according to the weight function, Eq.\ (\ref{tauWeight}), can be directly generated by 
the inverse-CDF (cumulative distribution function) method (i.e., without any rejected moves of the Metropolis algorithm). In this case,
the inverse CDF can furthermore be obtained analytically.\cite{Doran2019,Doran2020b} 

The redundant-walker algorithm\cite{Willow2013b} provides an essential speedup for the MC-MP2 algorithm and is invoked in the relativistic MC-MP2 also.
It propagates $m \gg 2$ electron pairs (walkers), $(\bm{r}^{[n]}_{1p},\bm{r}^{[n]}_{2p})$, $(1 \leq p \leq m)$, and permute their coordinates in all ${}_mC_2$ distinct ways when evaluating
the quotient of Eq.\ (\ref{MCMP2}), i.e.,
\begin{eqnarray}
E^{(2)} \approx \frac{1}{N} \sum_{n=1}^N I^\prime_n \equiv I_N
\end{eqnarray}
with
\begin{eqnarray}
 I^\prime_n = \frac{1}{{}_mC_2} \sum_{p < q}^{m} \frac{{f}(\bm{r}_{1p}^{[n]}, \bm{r}_{2p}^{[n]}, \bm{r}_{1q}^{[n]}, \bm{r}_{2q}^{[n]}, \tau^{[n]})}
{w(\bm{r}_{1p}^{[n]}, \bm{r}_{2p}^{[n]}) w(\bm{r}_{1q}^{[n]}, \bm{r}_{2q}^{[n]}) w(\tau^{[n]})}. \label{redundant}
\end{eqnarray}
It achieves a theoretical speedup by a factor of $(m-1)$ because the sampling is boosted by a factor of ${}_mC_2$ at an $m/2$-fold increase in the cost of walker propagation,
although the rate of speed up does not go up indefinitely with $m$ owing to saturation. 
See Ref.\ \onlinecite{Doran2016} for the observed performance of this algorithm.

The statistical uncertainty $\bar{\sigma}_N$ in $I_N$ is given by
\begin{eqnarray}\label{uncertainty}
\bar{\sigma}_N = \frac{\sigma_N}{\sqrt{N/N_\text{b}}}
\end{eqnarray}
where $\sigma_N$ is the the standard deviation concurrently accumulated (excluding the initial burn-in steps) as
\begin{eqnarray}
\sigma_N^2 = \frac{N_\text{b}}{N} \sum_{n=1}^{N/N_\text{b}} \left[ \frac{1}{N_\text{b}} \sum_{n^\prime=1}^{N_\text{b}}  I^\prime_{(n-1)N_\text{b} + n^\prime}- I_N \right]^2.
\end{eqnarray}
Here, $N_\text{b}$ is the block size in the Flyvbjerg--Petersen algorithm,\cite{Blocking} preventing underestimation of the standard deviation due to autocorrelation.

The whole algorithm can be easily and efficiently parallelized.\cite{Doran2016} Simply, each processor is tasked to perform its own independent MC integration using 
a distinct random number seed. It periodically records the number of MC steps, $I_N$, and $\bar{\sigma}_N$, with no interprocessor communications whatsoever.  
Only at the end of the calculation, weighted averages of the integrals from all surviving processors need to be taken. 
In this way, the algorithm is guaranteed to achieve near 100\% parallel efficiency, be fault-tolerant, be asynchronous, and be indefinitely restartable with a variable number of processors. 

\section{Numerical demonstrations}
For eight molecules, the relativistic four-component MP2 energy was obtained using the MC method:\ $\text{H}_2$ ($0.74\,\AA$), $\text{H}_2\text{O}$ ($r_{\text{OH}} = 0.96\,\AA$, $r_{\text{HH}} = 1.51\,\AA$), $\text{CuH}  $ ($1.54\,\AA$), $\text{AgH}$ ($1.70\,\AA$), $\text{AuH}$ ($1.57\,\AA$), $\text{Cu}_2  $ ($2.40\,\AA$), $\text{Ag}_2  $ ($2.68\,\AA$), and $\text{Au}_2$ ($2.59\,\AA$).  The basis functions used were contracted Gaussian-type orbitals:\cite{Yanai_2001-114, Yanai_2001-115}\ H [8s2p]/(3s2p), O [14s9p2d]/(5s4p2d), Cu [18s147d]/(8s6p3d), Ag [25s21p12d]/(11s9p4d), and Au [29s25p15d10f]/(11s8p5d3f). The weight functions are defined in Table \ref{basispam}.
The number of walkers [$m$ in Eq.\ (\ref{redundant})] used are given below.
The Flyvbjerg--Petersen block size of $N_\text{b} = 100$ was used.
A Dirac--Hartree--Fock calculation must precede the MC-MP2 step and was performed deterministically 
with the UTChem software,\cite{utchem,Yanai_2004} furnishing 
the MO coefficients and orbital energies. 
The benchmark deterministic relativistic MP2 calculations were also carried out with UTChem.
The MC-MP2 calculations were executed on an Intel core i9-10900K with 5.3 GHz and on the Cori (Haswell) supercomputer of NERSC.

\begin{table}
\caption{The weight-function parameters [Eq.\ (\ref{basisSet})].\label{basispam}}
\begin{ruledtabular}
 \begin{tabular}{lllll} 
Atom   & $c_1$  &  $\zeta_1$ & $c_2$  &  $\zeta_2$ \\ 
 \hline
 H &  0.25  & 0.06 &  0.15  &  0.6   \\
 O & 0.8 & 0.2  &  1.0  &  0.4   \\
 Cu  & 0.8 & 0.35 &  2.0  &  0.6   \\ 
 Ag & 0.1 & 0.1 &  0.8  &  0.6   \\
 Au & 0.05  & 0.6  &  4.0  &  0.8   \\ 
 \end{tabular} 
\end{ruledtabular}
\end{table}

\begin{table}
\caption{The relativistic second-order energies, $E^{(2)}$, and statistical uncertainties, $\bar{\sigma}$,
obtained by the MC-MP2 method after $10^5$ MC steps with eight electron-pair walkers ($m=8$). The benchmark $E^{(2)}$ obtained
by the deterministic MP2 method (as implemented in UTChem) are also shown.\label{results}}
\begin{ruledtabular}
\begin{tabular}{lccc}
& MP2 & \multicolumn{2}{c}{MC-MP2} \\
\cline{2-2} \cline{3-4}
Molecule    & $E^{(2)}\,/\, E_\text{h}$ & $E^{(2)}\,/\,E_\text{h}$ & $\bar{\sigma}\,/\,E_\text{h}$ \\
\hline
H$_2$   & $-0.0184$    &  $-0.0183$ & $0.0008$ \\
H$_2$O & $-0.2252$     & $-0.2248$ & $0.0106$ \\
CuH     & $-0.4618$    &  $-0.4634$ & $0.0198$  \\
AgH     & $-0.3212$    &  $-0.3243$ & $0.0152$ \\
AuH     & $-0.8424$    & $-0.8441$ & $0.0325$ \\
Cu$_2$  & $-0.8918$    & $-0.8931$ & $0.0371$\\
Ag$_2$  & $-0.6172$    & $-0.6146$ & $0.0235$ \\
Au$_2$  & $-3.0345$    & $-3.0316$ & $0.1432$ \\
\end{tabular} 
\end{ruledtabular}
\end{table}

 \begin{figure}
  \includegraphics[width=1\columnwidth]{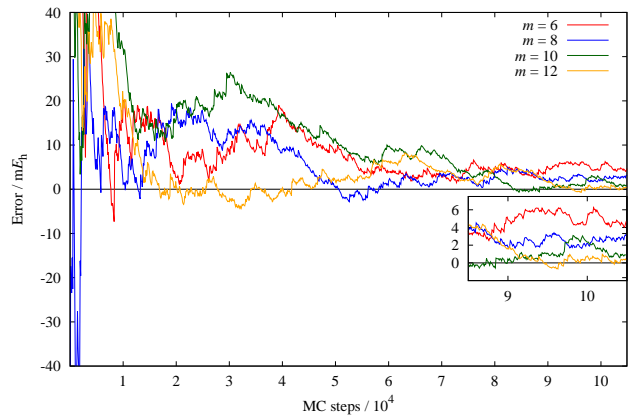}
  \caption{The error (in m$E_\text{h}$) in the relativistic MC-MP2 energy of AgH from the deterministic value as a function of the number of MC steps. Each for four traces 
  uses a different number ($m$) of electron-pair walkers.}\label{errors}
 \end{figure}
 
  \begin{figure}
      \includegraphics[width=1\columnwidth]{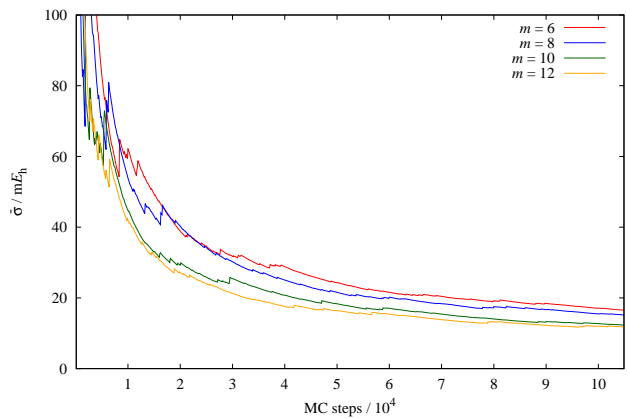}
   \caption{The statistical uncertainty $\bar{\sigma}$ (in $\text{m}E_\text{h}$) in the relativistic MC-MP2 energy of AgH  as a function of the number of MC steps. Each for four traces 
  uses a different number ($m$) of electron-pair walkers.}\label{vars}
 \end{figure}

In Table \ref{results} are presented the relativistic MC-MP2 energies ($E^{(2)}$) for the eight molecules mentioned above, using eight electron-pair walkers ($m=8$) 
at $10^5$ MC steps. The MC algorithm can reliably reproduce the deterministic results within the estimated statistical uncertainty $\bar{\sigma}$ even with
the relatively short MC runs.  
Although the MC-MP2 algorithm might most resemble the variational Monte Carlo method among the QMC variety, its energy is not bounded from below by the correct MP2 energy
and can be more negative than the latter.
The uncertainties systematically increase with the number of electrons. 
Figure \ref{errors} shows a typical convergence behavior of $E^{(2)}$ as a function of the number of MC steps for AgH using four different numbers of the electron-pair
walkers ($m = 6$, 8, 10, and 12). Figure \ref{vars} plots the corresponding statistical uncertainty ($\bar{\sigma}$). They attest to the acceleration of convergence with increasing
$m$ by the redundant-walker algorithm. 
 
 \begin{figure}
 \includegraphics[width=1\columnwidth]{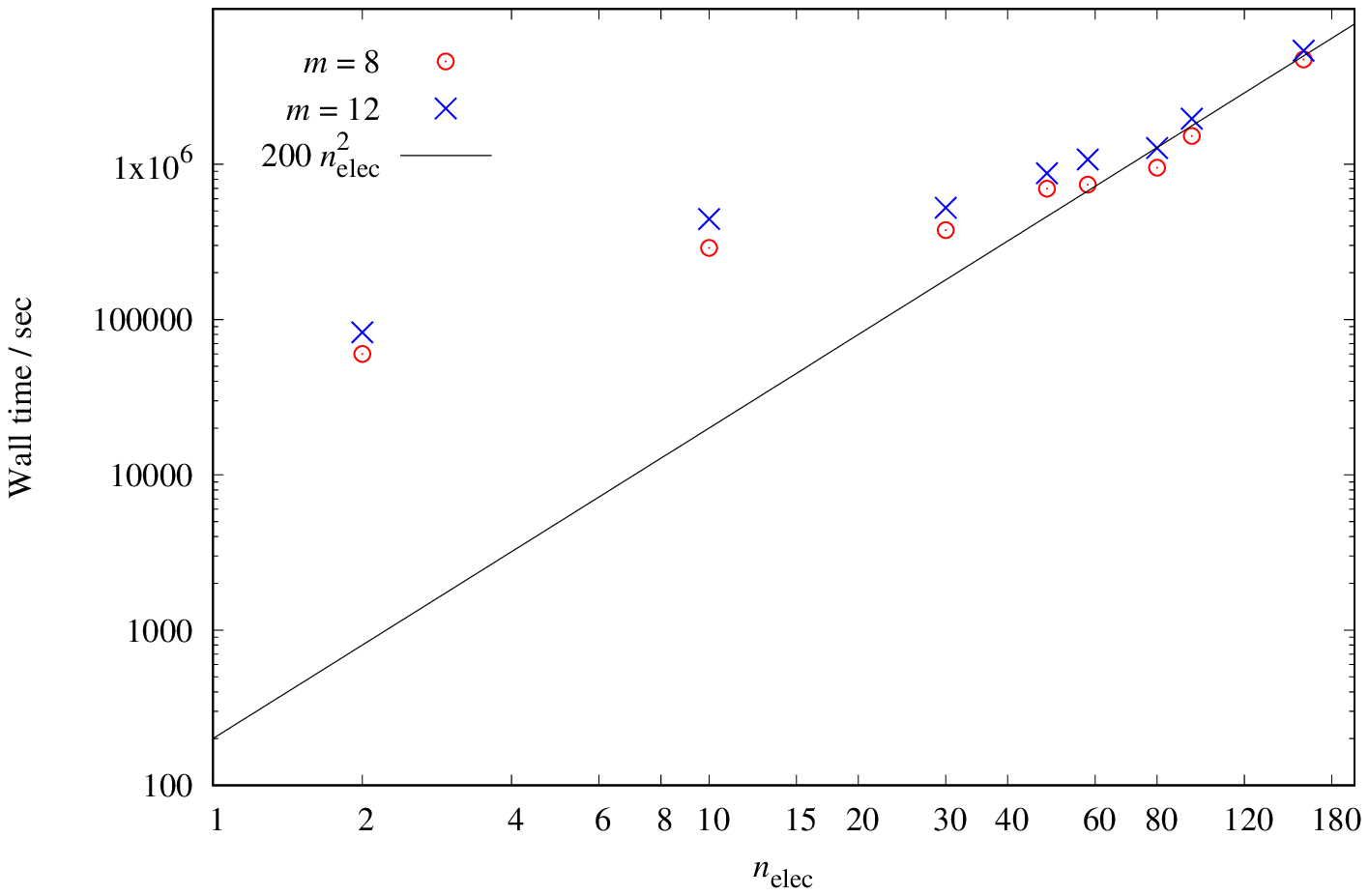}
  \caption{Wall time (in seconds) required for the relativistic MC-MP2 calculation to reach the relative statistical uncertainty ($\bar{\sigma}/|E^{(2)}|$) of 0.1 as a function of the number of electrons, $n_\text{ele}$. Two different values of the electron-pair walkers ($m=8$ and $12$) are considered. A line corresponding to quadratic cost scaling $O(n_\text{ele}^2)$
  is also plotted to guide the eyes. \label{ele}}
 \end{figure}
 
 \begin{figure}
 \includegraphics[width=1\columnwidth]{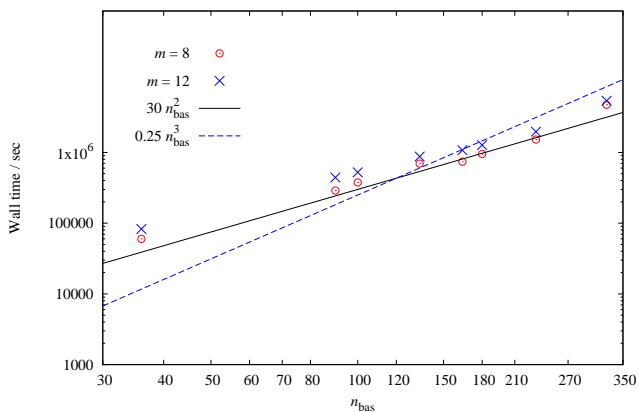}
  \caption{Wall time (in seconds) required for the relativistic MC-MP2 calculation to reach the relative statistical uncertainty ($\bar{\sigma}/|E^{(2)}|$) of 0.1 as a function of the number of basis functions, $n_\text{bas}$. Two different values of the electron-pair walkers ($m=8$ and $12$) are considered. Lines corresponding to quadratic and cubic cost scaling $O(n_\text{bas}^2)$ and $O(n_\text{bas}^3)$ are also plotted to guide the eyes. \label{spinor}}
 \end{figure} 

The computational cost of the MC algorithm can be measured by the wall time spent by the processor in a serial calculation to reach a given relative statistical error ($\bar{\sigma}/|E^{(2)}|$).\cite{Doran2016} The number of MC steps to reach the relative error can vary depending on molecules. 
These costs are plotted as a function of the number of electrons in Fig.\ \ref{ele} and of the number of basis functions in Fig.\ \ref{spinor}. In both plots, the cost function
are roughly quadratic with molecular size ($n_\text{ele}$ or $n_\text{bas}$) regardless of $m$ (the number of electron-pair walkers), although the last few data points 
may be taken to suggest asymptotic cubic scaling.
Note that nonrelativistic MC-MP2 is shown\cite{Doran2016} to display $O(n_\text{ele}^3)$
 cost scaling for spatially extended molecules, which is already a vast improvement over the $O(n_\text{ele}^5)$ scaling of the conventional, deterministic MP2.
It now appears possible that for crowded and large (but spatially compact) molecules with heavy elements, relativistic MC-MP2 can achieve even more favorable 
scaling of $O(n_\text{ele}^2)$. In the future, we shall further increase the number of electrons without spatially expanding the molecule to verify this scaling.
Our algorithmic analysis and observed timing data (not shown) clearly indicate that the cost per MC step is precisely quadratic with either $n_\text{ele}$ or $n_\text{bas}$. Therefore, for these crowded molecules,  the number of MC steps to reach the same relative uncertainty is roughly the same. This is also borne out in Table \ref{results}, where the relative uncertainties seem comparable 
after $10^5$ MC steps. It may be speculated that this is due to the fact that both $E^{(2)}$ and $\bar{\sigma}$ are dominated by similar numbers of valence spinors
regardless of molecular size.

 \begin{figure}
 \includegraphics[width=1\columnwidth]{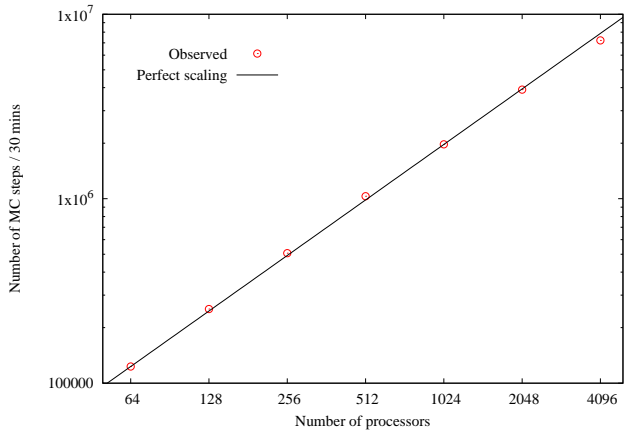}
  \caption{Number of MC steps taken in 30 minutes of wall clock time as a function of the number of processors executing the MC-MP2 calculation for Au$_2$ ($m=8$)
  on the Cori (Haswell) supercomputer of NERSC.\label{parallel}}
 \end{figure}

Figure \ref{parallel} plots the number of MC steps that can be carried out during the 30-minute wall clock time limit of a job queue on a NERSC supercomputer
as a function of the number of processors executing a 
MC-MP2 calculation for Au$_2$. Increasing the number of processors by a factor of 64 (from 64 to 4096 processors), we have observed a 59-fold speedup, i.e.,
strong scalability of 92\% for a fixed job size. This high parallel efficiency is expected because the algorithm is inherently parallel with no interprocessor communications. We can even let the calculation
be terminated by the job scheduler with no loss of data or waste of computational efforts. The data from various calculations can furthermore be concatenated to produce 
more converged energies. Some points in this plot are slightly above the line of perfect scalability relative to 64 processors, displaying small superscaling behavior. 


\section{Summary}

The Monte Carlo four-component relativistic MP2 algorithm has been developed. It is scalable 
with respect to both system size and computer size. For crowded, but spatially compact molecules with heavy elements, where the local-correlation approach
cannot apply, 
this stochastic algorithm displays tentative $O(n_\text{ele}^2)$ scaling and certainly no worse than $O(n_\text{ele}^3)$, which is a vast improvement over 
the $O(n_\text{ele}^5)$ scaling of the conventional MP2 algorithms. It can be implemented easily into a nearly perfectly scalable parallel algorithm with no interprocessor communications, which is also fault-tolerant, asynchronous, and indefinitely restartable.  We furthermore expect that it lends itself to facile extensions to  the Breit interactions, finite nucleus effects, and explicit correlation.\cite{tennoyamaki}

\section{Data Availability Statement}
A dataset is retrievable as a compressed file  at 
https:$//$hirata-lab.chemistry.illinois.edu$//$CruzJCP2022.zip.
It contains UTChem basis set and output files as well as spinor expansion coefficients for 
the eight molecules considered in this study. It also compiles the MC-MP2 output files for the energies and statistical uncertainties as a function of blocked MC steps for the eight molecules. The raw data used for plotting Figures 1 through 4 are also included.

\acknowledgments
J. C. Cruz thanks CONACYT for the scholarship 620190.
T. Yanai thanks JSPS KAKENHI (Grant No. 21H01881 and 21K18931) for financial support. 
S. Hirata thanks the Center for Scalable, Predictive methods for Excitation and Correlated phenomena (SPEC), which is funded by 
the U.S. Department of Energy, Office of Science, Office of Basic Energy Sciences, 
Chemical Sciences, Geosciences, and Biosciences Division, as a part of the Computational Chemical Sciences Program.
S. Hirata also thanks Grant No.\ DE-SC0006028 funded by the U.S. Department of Energy, Office of Science, Office of Basic Energy Sciences.
This research used resources of the National Energy Research Scientific Computing Center (NERSC), a U.S. Department of Energy Office of Science User Facility located at Lawrence Berkeley National Laboratory, operated under Contract No. DE-AC02-05CH11231 using NERSC award m3196 (2022).
The initial phase of this work was performed at Institute for Molecular Science, Okazaki, Japan.

%

\end{document}